\documentclass[12pt]{article}
\global\arraycolsep=2pt 
\usepackage{graphicx,amsmath,amssymb,euscript,psfrag,epsf,epsfig}
\usepackage{a4wide}
\usepackage{booktabs,dcolumn}
\usepackage{units}
\usepackage{feynmf}
\newcolumntype{d}[1]{D{.}{.}{#1}}
\makeatletter
\def\fmslash{\@ifnextchar[{\fmsl@sh}{\fmsl@sh[0mu]}}
\def\fmsl@sh[#1]#2{%
  \mathchoice
    {\@fmsl@sh\displaystyle{#1}{#2}}%
    {\@fmsl@sh\textstyle{#1}{#2}}%
    {\@fmsl@sh\scriptstyle{#1}{#2}}%
    {\@fmsl@sh\scriptscriptstyle{#1}{#2}}}
\def\@fmsl@sh#1#2#3{\m@th\ooalign{$\hfil#1\mkern#2/\hfil$\crcr$#1#3$}}
\makeatother
\begin{document}
\begin{titlepage}
\begin{flushright}
SI-HEP-2006-01 \\[0.2cm]
January 8, 2007
\end{flushright}

\vspace{1.2cm}
\begin{center}
\Large\bf\boldmath
Testing the left-handedness of the $b \to c$ transition 
\unboldmath 
 \end{center}

\vspace{0.5cm}
\begin{center}
{\sc B. M. Dassinger, R. Feger, T. Mannel} \\[0.1cm]
{\sf Theoretische Physik 1, Fachbereich Physik,
Universit\"at Siegen\\ D-57068 Siegen, Germany}
\end{center}

\vspace{0.8cm}
\begin{abstract}
\vspace{0.2cm}\noindent
We analyse the spin structure of inclusive semileptonic $b \to c$ transitions 
and the effects of non-standard model couplings on the rates and the spectra.
The calculation includes the ${\cal O} (\alpha_s)$ corrections as well as the
leading non-perturbative ones.
\end{abstract}

\end{titlepage}

\section{Introduction}
The enormous amount of data produced at the $B$ factories and the perspectives concerning future experiments with $b$ quarks will allow us to test the flavour non-diagonal sector of the standard model in a stringent way. However, aside from enough data on flavour changing decays this requires also a sufficient theoretical control in the calculation of the standard model predictions.

The sensitivity to ``new physics effects'' is expected to be largest in channels which do not have a big standard model contribution. These are in particular the loop induced flavour changing neutral current processes such as $b \to s$ transitions, which are suppressed by the GIM mechanism. Still there is also room for ``new physics effects'' in charged current processes, e.g. through the exchange of a heavy charged Higgs particle. 

The theoretical description of semi-leptonic decays has become very precise over the last couple of years and seems to be under good control \cite{MW}. In combination with the very large data sets from the $B$ factories this has led to precise determinations of of the CKM matrix elements $V_{cb}$ and $V_{ub}$, assuming that the standard model is correct \cite{BaBar,Belle}. 

In turn, these data may also serve to test the standard model. The charged currents will allow us to test aspects of the standard model different from the ones governed by loop induced flavour changing neutral current processes. In particular, the detailed data on semileptonic $B$ decays are sensitive to the helicity structure of the underlying quark transition, such as right-handed, scalar or tensor charged currents. 

The type of analysis we are suggesting here is well known in the context of leptonic muon and $\tau$ decays and is usually expressed as values or limits for the ``Michel parameters'' \cite{Michel1,Michel2}. In the present paper we analyse the $b \to c$ transition current along the same lines and discuss the sensitivity which can be obtained on new couplings on the basis of the precise data on spectral moments in inclusive semileptonic $B$ decays. 

The theoretical methods available for inclusive processes are in a mature state \cite{Bigiancient} and we shall compute the corresponding differential rates using the $1/m_b$ expansion \cite{MWold,TMold}. The state-of-the-art analysis within the standard model involves terms up to and including the terms of the order $1/m_b^3$ \cite{GK}, the ones of order $1/m_b^4$ \cite{DMT} and the next-to-leading QED and QCD radiative corrections to the leading (partonic) term \cite{Uraltsev}. In the present paper we perform the analysis for general currents at the appropriate level of precision.

In the next section we discuss the relevant operators using an effective field theory language. This allows us to identify the structure of the operators and will give us a qualitative argument on the sizes of the different contributions. In section 3 the OPE for the general interaction is performed, and in section 4 we discuss the radiative corrections to the leading term which is the partonic calculation. Finally we perform a numerical analysis of the effects of non-standard model interactions on the moments of semileptonic decay spectra and conclude.

\section{Effective Field Theory Description of ``New Physics''}
Since the standard model is the most general renormalizable theory with the observed particle spectrum and interactions, possible effects from physics beyond the standard model will show up as $SU(3) \times SU(2) \times U(1)$ invariant operators of higher dimension, which are suppressed by powers of the new-physics' scale. Focussing on the quark sector the symmetries of the standard model enforce that the leading operators are of dimension 6.

The list of relevant operators has been given in \cite{BuchmuellerWyler} and we shall use the notations of \cite{HansmannMannel}. The quark fields are grouped into

\begin{eqnarray}
Q_L &=& \left( \begin{array}{c} u_L \\ d_L \end{array} \right) , \quad 
               \left( \begin{array}{c} c_L \\ s_L \end{array} \right) , \quad 
               \left( \begin{array}{c} t_L \\ b_L \end{array} \right) \mbox{ for the left handed quarks and} \\
q_R &=& \left( \begin{array}{c} u_R \\ d_R \end{array} \right) , \quad
               \left( \begin{array}{c} c_R \\ s_R \end{array} \right) , \quad  
               \left( \begin{array}{c} t_R \\ b_R \end{array} \right) \mbox{ for the right handed quarks}
\end{eqnarray} 

where $Q_L$ are doublets under $SU(2)_L$ and $q_R$ are doublets under a (explicitly broken) $SU(2)_R$. The Higgs field and its charge conjugate are written as a $2 \times 2$ matrix
\begin{equation}
H = 1/\sqrt{2} \left( \begin{array}{cc} 
                      \phi_0 - i \chi_0 & \sqrt{2} \phi_+ \\
                      -\sqrt{2} \phi_-  & \phi_0 + i \chi \end{array} \right)
\end{equation}
transforming under $SU(2)_L \times SU(2)_R$. The potential of the Higgs fields leads to a vacuum expectation value (VEV) for the field $\phi_0$.

The $SU(2)_L \times SU(2)_R$ is broken by the Yukawa couplings and the weak hypercharge down to the standard model symmetry $SU(2)_L \times U(1)_Y$; however, keeping the notion of the larger $SU(2)_L \times SU(2)_R$ symmetry is useful for bookkeeping reasons.

The standard model is the most general renormalizable Lagrangian with an \linebreak $SU(2)_L\otimes U(1)_Y$ symmetry and the phenomenologically correct particle content.\linebreak This means that any effect from a high scale $\Lambda$ is suppressed by inverse powers of $\Lambda$, and --- in the sense of an effective field theory --- we are led to consider operators of dimension higher than 4. Hence, at scales of the order of the weak scale we may write the Lagrangian as an expansion in inverse powers of $\Lambda$ 
\begin{equation}
\mathcal{L}=\mathcal{L}_{4D}+\frac{1}{\Lambda} \mathcal{L}_{5D}+\frac{1}{\Lambda^2}\mathcal{L}_{6D}+...
\end{equation}
where the leading term is the standard model Lagrangian $\mathcal L_{SM} = \mathcal L_{4D}$. Furthermore, $\mathcal{L}_{5D}$, $\mathcal{L}_{6D}$ ... have to be invariant under the standard model symmetry $SU(2)_L \times U(1)_Y$. 

The next-to-leading terms in the $1/\Lambda$ expansion involve only dim-6 or higher operators, since for quarks there is no $SU(2)_L \otimes U(1)_Y$ invariant dim-5 operator. The list of possible operators is sizable and can be found e.g. in \cite{BuchmuellerWyler}. However, in the present paper we shall concentrate on semileptonic $B$ decays which restricts us to a smaller subgroup. It contains operators involving two quarks and gauge fields as well as four fermion operators, consisting of two quarks and two leptons. Since we know from semileptonic $\tau$ decays that a possible contribution from a four fermion operator at the weak scale is likely to be small, we shall focus on the operators with two quarks and gauge fields.

These two quark operators of dimension 6 may be classified according to the helicities of the quark fields, left-left (LL), right-right (RR) and left-right (LR). A full list of the relevant two quark operators can be found in \cite{HansmannMannel} where the following basis of operators has been suggested
\begin{eqnarray} \label{LL1}
O^{(1)}_{LL} &=& \bar{Q}_A\, \fmslash{L} G_{AB}^{(1)} \,Q_B, \\ \label{LL2}
O^{(2)}_{LL} &=& \bar{Q}_A\, \fmslash{L}_3G_{AB}^{(2)} \,Q_B
\end{eqnarray}
with
\begin{eqnarray}
L^\mu &=& H \left(i D^\mu H\right)^\dag + \left(i D^\mu H\right) H^\dag, \\
L^\mu_3 &=& H \tau_3 \left(i D^\mu H\right)^\dag +
            \left(i D^\mu H\right)\tau_3   H^\dag 
\end{eqnarray}
and all coupling matrices $G^{(i)}_{AB}$ being Hermitian. The terms proportional to $\tau_3$ have once again been included to break the custodial symmetry explicitly.

In the same spirit we define RR-operators
\begin{eqnarray}
O^{(1)}_{RR} &=& \bar{q}_A\, \fmslash{R}F_{AB}^{(1)} \,q_B \\
O^{(2)}_{RR} &=& \bar{q}_A\, \left\{ \tau_3,
            \fmslash{R}\right\} F_{AB}^{(2)} \,q_B \\ 
O^{(3)}_{RR} &=& i\bar{q}_A\, \left[  \tau_3,
        \fmslash{R}\right]  F_{AB}^{(3)}\,q_B \\
O^{(4)}_{RR} &=& \bar{q}_A\, \tau_3 \fmslash{R} \tau_3 F_{AB}^{(4)} \,q_B
\end{eqnarray}
with
\begin{eqnarray}
R^\mu &=& H^\dag \left(i D^\mu H\right)+ \left(i D^\mu H\right)^\dag H
\end{eqnarray}
and again Hermitian coupling matrices $F_{AB}^{(i)}$.

Using an odd number of Higgs fields we can construct invariant LR operators
\begin{eqnarray}
O^{(1)}_{LR} &=& \bar{Q}_A\, HH^\dag H \widehat{K}_{AB}^{(1)} \,q_B + h.c.
\label{hhh} \\
O^{(2)}_{LR} &=& \bar{Q}_A\, \left(\sigma_{\mu\nu}B^{\mu \nu} \right)H
\widehat{K}_{AB}^{(2)} \,q_B \, + h.c.\\
O^{(3)}_{LR} &=& \bar{Q}_A\,
\left(\sigma_{\mu\nu} W^{\mu\nu} \right) H\widehat{K}_{AB}^{(3)} \,q_B + h.c.\\
O^{(4)}_{LR} &=& \bar{Q}_A\,
\left(iD_\mu H\right) iD^\mu \widehat{K}_{AB}^{(4)} \,q_B + h.c.
\end{eqnarray}
with the coupling matrices 
\begin{equation}
\widehat{K}_{AB}^{(i)} = {K}^{(i)}_{AB} + \tau_3 {K}_{AB}^{(i)\prime} \, .
\end{equation}

After spontaneous symmetry breaking we want to consider the contributions of these operators to the charged current $b \to c$ transitions. The relevant interactions are
\begin{eqnarray}
O^{(1)}_{LL} &=& \frac{v^2 g }{\sqrt2} G_{cb}^{(1)} V_{cb} \,
           \bar{c}  \fmslash{W}^+ P_- b \\
O^{(1)}_{RR} &=& \frac{v^2 g }{\sqrt2} F_{cb}^{(1)} V_{cb}\,
           \bar{c}\fmslash{W}^+ P_+ b \\
O^{(3)}_{RR} &=&   \frac{v^2 g }{\sqrt2} 2iF_{cb}^{(3)} V_{cb}\,
           \bar{c}\fmslash{W}^+ P_+ b \\
O^{(4)}_{RR} &=& -  \frac{v^2 g }{\sqrt2} F_{cb}^{(4)}V_{cb}\, 
           \bar{c}\fmslash{W}^+ P_+ b \\
O^{(3)}_{LR} &=&  \frac{vg }{2}V_{cb} \,\bar{c} \sigma^{\mu \nu} \left\lbrace 
\partial_\mu W_\nu^+ \widetilde{K}_{cb}^{(3)} P_+ + \partial_\nu W_\mu^+ \widetilde{K}_{cb}^{\dagger(3)} P_-
\right\rbrace b \\
O^{(4)}_{LR} &=&   \frac{vg}{2}V_{cb}\, \bar{c} \left\lbrace 
W_\mu^+ i\partial^\mu \widetilde{K}_{cb}^{(4)} P_+ + W_\mu^+ i\partial^\mu \widetilde{K}_{cb}^{\dagger(4)} P_-
\right\rbrace b
\end{eqnarray}
containing the couplings $\widetilde{K}_{cb}^{(i)}= K_{cb}^{(i)} - K_{cb}^{\prime(i)}$ and the projectors $P_{\pm}=(1\pm \gamma_5)/2$.

Including also the contribution from the standard model 
\begin{equation}
\mathcal{L}_{SM} =  \frac{g}{\sqrt2} V_{cb} \, \bar{c}\fmslash{W}^+ P_- b \, , 
\end{equation}
we get the generalized interaction as 
\begin{equation}\label{SpecialBdecayL}
\begin{split}
\mathcal{L} &=\frac{g}{\sqrt2}\left(1+\frac{v^2}{\Lambda^2} G_{cb}^{(1)}\right)V_{cb}\,  
                     \bar{c} \fmslash{W}^+ P_- b \\
& + \frac{v^2 g}{\sqrt2}\frac{1}{\Lambda^2}\left( F_{cb}^{(1)} - F_{cb}^{(4)} +2iF_{cb}^{(3)}\right)V_{cb}\, \bar{c} \fmslash{W}^+ P_+ b\\
& +  \frac{vg}{2}\frac{1}{\Lambda^2}V_{cb}\, \bar{c} \left\lbrace 
W_\mu^+ i\partial^\mu \widetilde{K}_{cb}^{(4)} P_+ + W_\mu^+ i\partial^\mu \widetilde{K}_{cb}^{\dagger(4)} P_-
\right\rbrace  b\\
& +\frac{vg}{2}\frac{1}{\Lambda^2}V_{cb}\, \bar{c}  \sigma^{\mu \nu} \left\lbrace 
\partial_\mu W_\nu^+ \widetilde{K}_{cb}^{(3)} P_+ + \partial_\nu W_\mu^+ \widetilde{K}_{cb}^{\dagger(3)} P_-
\right\rbrace b .
\end{split}
\end{equation}

Going to hadronic energy scales we also integrate out the weak bosons. Since possible new physics contributions to the leptonic sector are strongly constrained, we shall use here only the standard model piece. Integrating out the weak bosons amounts to the replacement 
\begin{equation}
W_\mu^- = \frac{4 G_F}{\sqrt2}\bar e \gamma_\mu P_- \nu_e.
\end{equation}
from which we obtain the effective Hamiltonian
\begin{equation}\label{Hamiltonian2}
\mathcal{H}_{eff} = \frac{4G_F V_{cb}}{\sqrt2} J_{q,\mu} J_l^\mu,
\end{equation}
where $J_l^\mu = \bar{e}\, \gamma^\mu P_-\, \nu_e$ is the usual leptonic current and $J_{h,\mu}$ is the generalized hadronic $b \to c$ current which is given by
\begin{eqnarray} \label{EnhancedGamma}
J_{h,\mu}  &=& c_L \, \, \bar{c} \gamma_\mu P_- b  
                      + c_R  \, \, \bar{c} \gamma_\mu P_+ b
                      + g_L \, \,\bar{c} \frac{i D_\mu}{m_b} P_- b 
                      + g_R \, \, \bar{c} \frac{i D_\mu}{m_b} P_+ b \\
&& +  d_L \, \, \frac{i \partial^\nu}{m_b}  ( \bar{c}  i \sigma_{\mu \nu} P_- b)
      + d_R \, \, \frac{i \partial^\nu}{m_b}  ( \bar{c}  i \sigma_{\mu \nu} P_+ b) \, ,
      \nonumber 
\end{eqnarray}
where we have renamed the coupling constants to streamline the notation, $P_\pm$ denotes the projector on positive/negative chirality and $D_\mu$ is the QCD covariant derivative. Note that the term proportional to $c_L$ contains the standard model contribution.

In principle one could have written (\ref{EnhancedGamma}) from the start, but the derivation via an effective field theory approach allows us to consider the expected orders of magnitudes of the different contributions. Note that the derivatives acting on the fields in (\ref{SpecialBdecayL}) will yield momenta of the particles involved in the decay (see (\ref{EnhancedGamma})), and hence the typical scale which has to be assigned to these derivatives is the mass of the $b$ quark. Furthermore, we have already included the corresponding inverse powers of $m_b$ in the definitions of the couplings $g_{L/R}$ and $d_{L/R}$.

Comparing (\ref{EnhancedGamma}) with (\ref{SpecialBdecayL}) we get an estimate for the orders of magnitudes
\begin{equation}\label{Groessenordnungen}
	c_L \propto 1; \hspace{1cm} c_R \propto \frac{v^2}{\Lambda^2}; \hspace{1cm} d_{R/L}\propto\frac{v\, m_b}{\Lambda^2}; \hspace{1cm} g_{R/L}\propto\frac{v\, m_b}{\Lambda^2};
\end{equation}
where a possible overall loop factor $1/(16 \pi^2)$ has been omitted. 

We note further that in minimal flavour violating scenarios \cite{MFV} any occurrence of a right handed quark is related to a helicity flip and hence mass factors occur. While the above estimates remain the same for $g_{R/L}$ and $d_{R/L}$, the estimate for $c_R$ contains a strong additional suppression factor of $m_b m_c /v^2$. 

The additional factor $m_b / \Lambda$ in $g_{L/R}$ and $d_{L/R}$ reflects the fact that in order to obtain a helicity flip one has to have an additional Yukawa coupling, making these contributions small compared to the helicity-conserving ones.

In the (differential) rate the relevant contributions are the interference terms with the standard model piece. A coupling between left and right handed contributions requires to flip the helicity of the final state $c$ quark and hence an additional factor $m_c / m_b$ will appear in these contributions. 

Thus for the rates we obtain additional contributions of the orders
$c_L c_R \sim (m_c /m_b) v^2 / \Lambda^2$, 
$c_L d_L \sim c_L g_L \sim (m_b  v) / \Lambda^2$ and
$c_L d_R \sim c_L g_R \sim (m_c /m_b) (m_b  v) / \Lambda^2$,
while all other contributions will be too small to be relevant here.

\section{OPE for the Differential Inclusive Rate}
The calculation of the differential rate proceeds along the same lines as the corresponding calculation in the standard model. The starting point is a correlator of the two hadronic currents 

\begin{equation} \label{hadcorr}
T_{\mu \nu} = \int d^4 x \, e^{-ix(m_b v-q)} \langle B(p) |\bar{b}_v (x) \Gamma_\mu c(x)
\, \bar{c}(0) \Gamma_\nu^\dagger b_v(0) | B(p) \rangle,
\end{equation}

where $\Gamma$ is the Dirac matrix corresponding to (\ref{EnhancedGamma}), $v = p/M_B$ is the four velocity of the decaying $B$ meson and $q$ is the momentum transferred to the leptons. 

This correlator may be decomposed into scalar form factors according to 
\begin{equation}
T_{\mu \nu} = - g_{\mu \nu} T_1 + v_\mu v_\nu T_2 - i \epsilon_{\mu \nu \alpha \beta} 
                       v^\alpha q^\beta T_3 + q_\mu q_\nu T_4 
                       + (q_\mu v_\nu + v_\mu q_\nu ) T_5.
\end{equation}

Contracting the imaginary part of $T_{\mu \nu}$ with the tensor $L_{\mu \nu}$ 
obtained from the leptonic currents one obtains the differential decay rate 
\begin{equation}
d \Gamma = \frac{G_F^2}{4 M_B}  {\rm Im} T_{\mu \nu} L^{\mu \nu} d \phi_{\rm PS} 
\end{equation}
where $d \phi_{\rm PS}$ is the corresponding phase space element. 

As discussed above, we shall use the standard model expression for the leptonic side of the process. Using the charged lepton energy $E_\ell$, the neutrino energy $T$ and the leptonic invariant mass $S$ as independent variables one obtains for the triply differential rate 
\begin{equation}
\frac{d^3 \Gamma}{dE_\ell \, dT dS^2} = \frac{G_F^2 m_b}{4 \pi^3} 
\left[2 m_b S^2 \, {\rm Im} T_1 - m_b (S^2 - 4 E_\ell T) \, {\rm Im} T_2 
        - 2 S^2m_b (T-E_\ell)\,  {\rm Im} T_3 \right]
\end{equation}
In the following calculation we shall adopt the notations of \cite{Uraltsev}.

\subsection{Tree level expansion}
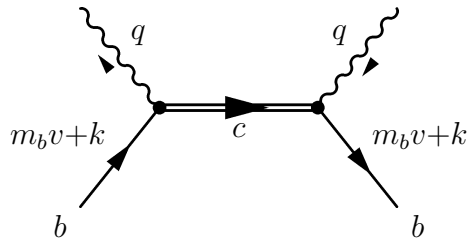
\begin{figure}[!h]
	\begin{center}
		\begin{fmffile}{bcb}
			\fmfcmd{%
				vardef middir(expr p,ang) =
				dir(angle direction length(p)/2 of p + ang)
				enddef;
				style_def arrow_left expr p =
				shrink(.7);
				cfill(arrow p
					shifted(4thick*middir(p,90)));
				endshrink
				enddef;
				style_def arrow_right expr p =
				shrink(.7);
				cfill(arrow p
					shifted (4thick*middir(p,-90)));
				endshrink
				enddef;
			}
			\begin{fmfgraph*}(150,75)
				\fmfleft{i1,i2}\fmflabel{$b$}{i1}
				\fmfright{o1,o2}\fmflabel{$b$}{o1}
				\fmf{boson,label=$q$,label.side=left}{i2,v1}
				\fmf{fermion,label=$\displaystyle m_bv{+}k$,label.side=left}{i1,v1}
				\fmf{double_arrow,label=$c$}{v1,v2}
				\fmf{boson,label=$q$,label.side=left}{v2,o2}
				\fmf{fermion,label=$\displaystyle m_bv{+}k$,label.side=left}{v2,o1}
				\fmfdot{v1,v2}
				\fmffreeze
				\fmf{arrow_left}{o2,v2}
				\fmf{arrow_left}{v1,i2}
			\end{fmfgraph*}
		\end{fmffile}
	\end{center}
	\caption{Feynman diagrams for the tree-level expansion in $1/m_b$}
    \label{fig:feyn:Tree}
\end{figure}
The tree-level expansion in $1/m_b$ is most easily set up along the lines described in \cite{DMT}. In the Feynman diagram shown in fig.~\ref{fig:feyn:Tree}, the double line denotes the propagator of a charm quark which propagating in the background field of the soft gluons of the $B$ meson
\begin{equation} \label{BGFprop}
	S_{\rm BGF} = \frac{-i}{\fmslash{Q}+ i \fmslash{D}-m_c} 
\end{equation}
where $Q=m_b v -q$ and $D$ denotes the covariant derivative with respect to the background gluon field. 

The OPE of the forward scattering amplitude is obtained by multiplying (\ref{BGFprop}) by the appropriate Dirac structures for the currents (see (\ref{EnhancedGamma}) and by expanding (\ref{BGFprop}) to third order in $iD$
\begin{equation} \label{ExpBGFprop}
i S_{\rm BGF} = \frac{1}{\fmslash{Q}-m_c} - 
\frac{1}{\fmslash{Q}-m_c} ( i \fmslash{D}) \frac{1}{\fmslash{Q}-m_c}
+ \frac{1}{\fmslash{Q}-m_c} ( i \fmslash{D}) \frac{1}{\fmslash{Q}-m_c}
  ( i \fmslash{D}) \frac{1}{\fmslash{Q}-m_c} + \cdots
\end{equation}
keeping track of the ordering of the covariant derivatives.

The remaining task is to evaluate the matrix elements of operators of the form
\[b_v (iD_{\mu_1}) ....  (iD_{\mu_n}) b_v \]
where $b_v$ is the full QCD field. This is done most conveniently in a recursive fashion as described in \cite{DMT}.

Expanding to $1/m_b^2$ requires the two well known matrix elements of dimension 5, which are the kinetic energy parameter $\mu_\pi$ and the chromomagnetic moment $\mu_G$
\begin{eqnarray}
2 M_B \mu_\pi^2 &=& -
\langle B(p) | \bar{b}_v (i D)^2 b_v | B(p) \rangle \\
2 M_B \mu_G^2  &=& 
\langle B(p) | \bar{b}_v (i D_\mu) (i D_\nu) 
(-i \sigma^{\mu \nu}  b_v | B(p) \rangle 
\end{eqnarray}

The dim-4 matrix elements contain only a single derivative, while the dim-3 contributions have no derivative. They can be expressed in terms of $\mu_\pi$, $\mu_G$ up to ${\cal O} (1/m_b^2)$ accuracy. The relevant general matrix elements are
\cite{DMT}
\begin{eqnarray}
&& \langle B(p)\, | \bar{b}_v (iD^\rho) (iD^\sigma) b_v\, |\, B(p)\, \rangle 
=  - 2 M_B \left[ \frac{ 1 }{6} P_+  \left(g^{\rho \sigma }- v^{\rho } v^{\sigma }\right) \mu_{\pi }^2 
 - \frac{ 1 }{12} P_+ (- i \sigma ^{\rho \sigma}) P_+ \mu_{G}^2 \right] ,
 \nonumber \\ 
&& \langle B(p) | \bar{b}_v (iD^\rho) b_v | B(p) \rangle = -\frac{M_B}{2 m_b} P_+  
\bigg\lbrace v^\rho  (\mu^2_G-\mu_\pi^2) \bigg \rbrace  
+ \frac{M_B}{6 m_b} \bigg\lbrace (\gamma^\rho -v^\rho \fmslash{v}) 
	    (\mu^2_G-\mu_\pi^2) \bigg \rbrace,   \nonumber \\
&& \langle B(p) | \bar{b}_v b_v | B(p) \rangle =  P_+\,\, M_B + \frac{M_B}{4m_b^2}(\mu^2_G-\mu_\pi^2),
\end{eqnarray}
and the result to order $1/m_b^2$ is obtained using the expansion (\ref{ExpBGFprop}) and the general vertices (\ref{EnhancedGamma}).

In this way we evaluate the scalar components of the correlator $T_{\mu \nu}$ given in (\ref{hadcorr}). Although it is straightforward to compute all contributions we shall give here only the ones which can be expected to be sizable, these are the interferences with the standard model contribution. We write 
\begin{equation}
T_i = T_i^{\rm c_L c_L} + T_i^{\rm c_L c_R} + T_i^{\rm c_L g_L} + T_i^{\rm c_L g_R}
+ T_i^{\rm c_L d_L} + T_i^{\rm c_L d_R},
\end{equation}
where all scalar components have an expansion in $1/m_b$ according to
\begin{equation}
T_i^{\rm c_L c_L} = T_{i,0}^{\rm c_L c_L} + 
                                T_{i,2}^{\rm c_L c_L} + T_{i,3}^{\rm c_L c_L} + ...
\end{equation} 

For the standard model contribution this expansion has be performed already to $1/m_b^4$ \cite{DMT}, while for the new physics contributions it is sufficient to retain the tree level terms, since the nonperturbative corrections at least to the low moments are tiny \cite{dassing,feger}. More relevant are the QCD radiative corrections which we shall consider in the next section.

\subsection{QCD Radiative Corrections} 
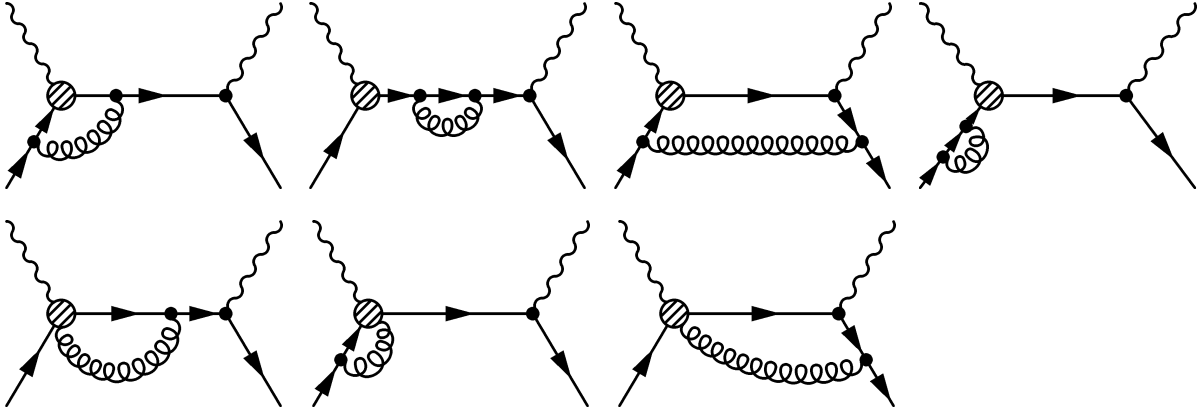
\begin{figure}[!h]
	\begin{fmffile}{alphas}
		\newenvironment{bcb}
			{\begin{minipage}{3.8cm}
				\begin{fmfgraph*}(130,70)
					\fmfset{curly_len}{2mm}
					\fmfleft{i1,i2}
					\fmfright{o1,o2}
					\fmf{boson}{i2,v2}
					\fmf{boson}{v5,o2}
					\fmf{phantom,tension=2}{i1,v1,v2,v3,v4,v5,v6,o1}
					\fmfdot{v5}
					\fmfblob{.08w}{v2}
					\fmffreeze
			}
			{	\end{fmfgraph*}
			 \end{minipage}
			}
		\newenvironment{bcbren}
			{\begin{minipage}{3.8cm}
				\begin{fmfgraph*}(130,70)
					\fmfset{curly_len}{2mm}
					\fmfleft{i1,i2}
					\fmfright{o1,o2}
					\fmf{boson}{i2,v3}
					\fmf{boson}{v6,o2}
					\fmf{phantom,tension=3}{i1,v1,v2,v3,v4,v5,v6,v7,v8,o1}
					\fmfdot{v6}
					\fmfblob{.08w}{v3}
					\fmffreeze
			}
			{	\end{fmfgraph*}
			 \end{minipage}
			}
		\begin{bcb}
			\fmf{fermion}{i1,v1,v2,v5,o1}
			\fmf{gluon,right=.5}{v1,v3}
			\fmfdot{v1,v3}
		\end{bcb}
		\begin{bcb}
			\fmf{fermion}{i1,v2,v3,v4,v5,o1}
			\fmf{gluon,right}{v3,v4}
			\fmfdot{v3,v4}
		\end{bcb}
		\begin{bcb}
			\fmf{fermion}{i1,v1,v2,v5,v6,o1}
			\fmf{gluon}{v1,v6}
			\fmfdot{v1,v6}
		\end{bcb}
		\begin{bcbren}
			\fmf{fermion}{i1,v1,v2,v3,v6,o1}
			\fmf{gluon,right}{v1,v2}
			\fmfdot{v1,v2}
		\end{bcbren}
		\vspace{4mm}

		\begin{bcb}
			\fmf{fermion}{i1,v2,v4,v5,o1}
			\fmf{gluon,right}{v2,v4}
			\fmfdot{v4}
		\end{bcb}
		\begin{bcb}
			\fmf{fermion}{i1,v1,v2,v5,o1}
			\fmf{gluon,right}{v1,v2}
			\fmfdot{v1}
		\end{bcb}
		\begin{bcb}
			\fmf{fermion}{i1,v2,v5,v6,o1}
			\fmf{gluon,right=.3}{v2,v6}
			\fmfdot{v6}
		\end{bcb}
	\end{fmffile}
	\caption{Feynman diagrams for the ${\cal O} (\alpha_s)$ radiative corrections.}
	\label{fig:feyn:Alphas}
\end{figure}

The main corrections to the non-standard model contributions is due to QCD radiative effects, since at the current level of precision the non-perturbative corrections are too small to be relevant. Hence we consider now the order $\alpha_s$ radiative corrections, which are computed by evaluating the Feynman diagrams shown in fig.~\ref{fig:feyn:Alphas}. Note that the scalar current involves also an effective quark-quark-gluon-boson vertex in order to maintain QCD gauge invariance. Taking the imaginary part at the end implements the various cuts of the diagrams in fig.~\ref{fig:feyn:Alphas} corresponding to real and virtual radiation.

The standard model contribution involves only left handed currents and has been computed already long ago in \cite{veryold1,veryold2}. The left handed current is a dim-3 operator and has a vanishing anomalous dimension; hence the standard model calculation for the sum of the real and the virtual radiation yields an ultraviolet finite result.

As discussed above, the non-standard model operators correspond to dim-6 operators, but after spontaneous symmetry breaking the operators may be treated as dim-3 or dim-4-operators as far as the QCD corrections are concerned.

For the right handed current the same argument holds as for the left handed current. It is also a dim-3 operator and is conserved; thus the anomalous dimension vanishes and the sum of the real and the virtual QCD corrections is ultraviolet finite.

The tensor and the scalar operators connect different helicities and thus they contain only a single power of the weak VEV and in addition one derivative. Consequently, from the QCD point-of-view they are dim-4 operators which renormalize in QCD.

However, we are interested only in the QCD effects on the shape of the spectra and hence the virtual corrections do not play a role, since they only renormalize the tree level result. In particular, the partonic mass moments $\langle (p_X^2-m_c^2)^n \rangle$ (with $n>0$) do not have a tree level contribution (and hence no contribution from virtual QCD corrections), since the tree level rate is proportional to $\delta (p_X^2-m_c^2)$. Hence we need to compute only the contributions to the real radiation and can ignore - at least for this purpose - the question of the renormalization of the helicity-changing operators.

\section{Results and Discussion} 
The new physics contributions will have an impact on the shapes of the spectra of the hadronic and leptonic energies as well as on the spectrum of the hadronic invariant mass.

It is instructive to first take a look at the tree level results for the charged lepton energy spectrum. Using the variable $y = 2 E_\ell /m_b$ we  write the different contributions as
\begin{equation}\label{lskducnem}
\frac{\text{d}\Gamma}{\text{d}y}=\sum_i \frac{\text{d}\Gamma^{(i)}}{\text{d}y}
\quad , \, i=c_L c_L, \, c_L c_R, \, c_L d_L, \, c_L d_R, \, c_L g_L, \, c_L g_R
\end{equation}
where the superscript denotes the different contributions according to the 
coupling constants as defined in (\ref{EnhancedGamma}). The term proportional 
to $c_L c_L$ contains the well known standard-model contribution
\begin{equation} 
\frac{\text{d}\Gamma^{c_Lc_L}}{\text{d}y} =
\frac{G_F^2|V_{cb}|^2 m_b^5}{192  \pi^3}
\left[ 2 y^2 (3 - 2y) - 6 y^2 \rho - \frac{6 y^2 \rho^2 }{(1-y)^2}
    + \frac{2 y^2 (3-y) \rho^3}{(1-y)^3} \right],
\end{equation} 
while the contribution from a right handed current is given by
\begin{equation}
\frac{\text{d}\Gamma^{c_Lc_R}}{\text{d}y} = - 
\frac{G_F^2|V_{cb}|^2 m_b^3}{192\,\pi^3} \sqrt{\rho}
\left[ 12 y^2 - \frac{24 y^2 \rho}{1-y} + \frac{12 y^2 \rho^2 }{(1-y)^2} \right] \, .
\end{equation}
The tree level results involving the tensor currents is 
\begin{equation}
\frac{\text{d}\Gamma^{c_L d_R}}{\text{d}y} = -
\frac{G_F^2|V_{cb}|^2 m_b^5}{192 \pi^3}
\left[ 4 y^3 - \frac{12 y^3 \rho^2}{(1-y)^2} + \frac{8 y^3 \rho^3}{(1-y)^3} \right] 
\end{equation}
and
\begin{equation}
\frac{\text{d}\Gamma^{c_Ld_L}}{\text{d}y}= - 
\frac{G_F^2|V_{cb}|^2 m_b^5}{192 \pi^3} \sqrt\rho
\left[ \frac{4 y^2 (3-y) (y-1+\rho)^3}{(1-y)^3} \right]
\end{equation}
Likewise, the tree contributions involving the scalar current are
\begin{equation}
\frac{\text{d}\Gamma^{c_Lg_L}}{\text{d}y} = 
\frac{G_F^2|V_{cb}|^2m_b^5}{192 \pi^3} \sqrt\rho
\left[ \frac{6 y^2 (y-1+\rho)^2} {1-y} \right] = \sqrt{\rho} \,\, 
\frac{\text{d}\Gamma^{c_Lg_R}}{\text{d}y}.
\end{equation}

\begin{figure}
	\begin{center}
		\includegraphics{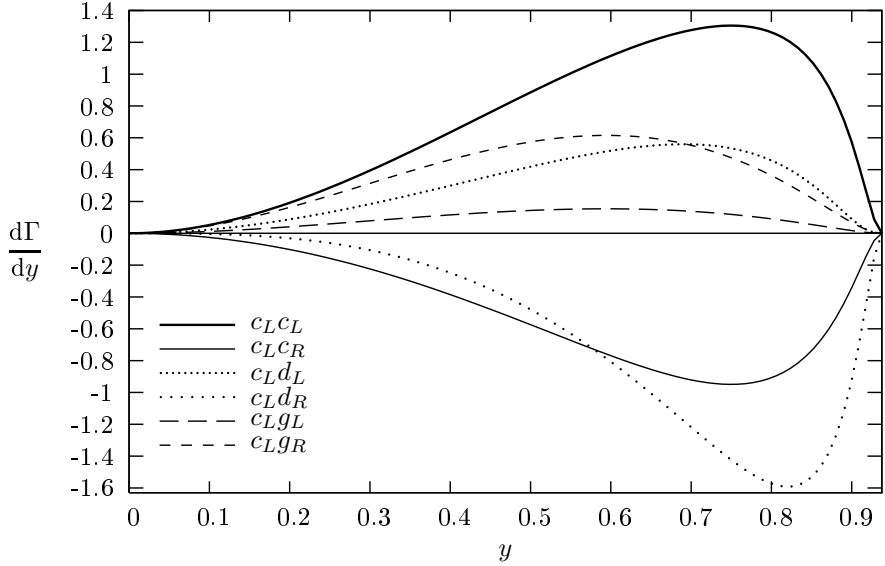}
	\end{center}
	\vspace{-3mm}
	\caption{Effect of the additional contributions on the lepton energy spectrum, normalized to $G_F^2|V_{cb}|^2m_b^5/(192\pi^3)$.}
	\label{fig:spectra}
\end{figure}

In fig.~\ref{fig:spectra} we show the effect of the additional contributions in the hadronic current at tree level. This plot already shows that a contribution of a different helicity has a large impact on the shape of the spectrum.

It is well known that the moments are sufficiently inclusive and can be calculated reliably in the $1/m_b$ expansion, even if a (not too high) cut on the lepton energy is applied. In the following tables we give the results for various moments without an energy cut for the charged lepton energy $E_l$ and with a cut of 1 GeV for this quantity. In the tables we list the results for
\begin{equation}
L_n = \frac{1}{\Gamma_0} 
\int_{E_{\rm cut}} d E_\ell \, E_\ell^n \, \frac{d \Gamma}{d E_\ell}
\end{equation}
and 
\begin{equation}
H_{ij} = \frac{1}{\Gamma_0}
\int_{E_{\rm cut}} d E_\ell \, \int dE_{\rm had} \, dM_{\rm had}^2 \,
 (M_{\rm had}^2-m_c^2)^i \, E_{\rm had}^j \,
\frac{d^3 \Gamma}{d E_\ell \, dE_{\rm had} d M_{\rm had}^2} 
\end{equation}
with 
\begin{equation}
\Gamma_0 = \frac{G_F^2|V_{cb}|^2 m_b^5}{192 \pi^3}
\left[1-8\rho -12 \rho^2 \ln \rho + 8 \rho^3 - \rho^4 \right]
\end{equation}

The entries in the tables contain the coefficients corresponding to the expansion (\ref{lskducnem}) of the differential rate
\begin{eqnarray*}
L_n &=& c_L^2 L_n^{(c_L c_L)} + c_L c_R L_n^{(c_L c_R)}
            + c_L d_L L_n^{(c_L d_L)}  + c_L d_R L_n^{(c_L d_R)}
             + c_L g_L L_n^{(c_L g_L)} + c_L g_R L_n^{(c_L g_R)} \\  
H_{ij}  &=& c_L^2 H_{ij} ^{(c_L c_L)} + c_L c_R H_{ij} ^{(c_L c_R)} 
             + c_L d_L H_{ij} ^{(c_L d_L)}  + c_L d_R H_{ij}^{(c_L d_R)} 
             + c_L g_L H_{ij} ^{(c_L g_L)} + c_L g_R H_{ij} ^{(c_L g_R)}
\end{eqnarray*}
For the following numerical estimates we use $m_b = 4.6$ GeV and $m_c = 1.15$ GeV, resulting in $\rho = m_c^2 / m_b^2 = 0.0625$.

Table~\ref{tab:LTree} contains the results for the tree level contributions for the moments of the lepton energy spectrum without cut, table~\ref{tab:LTreeCut} contains the same quantities including a lepton energy cut.

\begin{table}
	\begin{center}
	\begin{tabular}{c|d6d6d6d6d6d6}\toprule
		n & \multicolumn{1}{c}{$c_l^2$} & \multicolumn{1}{c}{$\:\ c_lc_r$} & \multicolumn{1}{c}{$c_ld_l$} & \multicolumn{1}{c}{$\:\ c_ld_r$} & \multicolumn{1}{c}{$c_lg_l$} & \multicolumn{1}{c}{$c_lg_r$} \\
		\midrule
		0 & 1.000000 & -0.668451 & 0.418451 & -0.832887 & 0.125000 & 0.500000 \\
		1 & 0.307202 & -0.209226 & 0.123322 & -0.291310 & 0.033849 & 0.135394 \\
		2 & 0.103000 & -0.070764 & 0.039683 & -0.106637 & 0.010209 & 0.040836 \\
		3 & 0.036524 & -0.025173 & 0.013523 & -0.040307 & 0.003306 & 0.013223 \\
		\bottomrule
	\end{tabular}
	\end{center}
	\caption{Tree level coefficients of the leptonic moments without $E_l$ cuts.}
	\label{tab:LTree}
\end{table}

\begin{table}
	\begin{center}
	\begin{tabular}{c|d4d4d4d4d4d4}\toprule
		n & \multicolumn{1}{c}{$c_l^2$} & \multicolumn{1}{c}{$\:\ c_lc_r$} & \multicolumn{1}{c}{$c_ld_l$} & \multicolumn{1}{c}{$\:\ c_ld_r$} & \multicolumn{1}{c}{$c_lg_l$} & \multicolumn{1}{c}{$c_lg_r$} \\
		\midrule
		0 & 0.8151 & -0.5619 & 0.3322 & -0.7780 & 0.0887 & 0.3559 \\
		1 & 0.2775 & -0.1921 & 0.1095 & -0.2816 & 0.0281 & 0.1128 \\
		2 & 0.0980 & -0.0677 & 0.0374 & -0.1049 & 0.0092 & 0.0370 \\
		3 & 0.0356 & -0.0246 & 0.0131 & -0.0400 & 0.0031 & 0.0126 \\
		\bottomrule
	\end{tabular}
	\end{center}
	\caption{Tree level coefficients of the leptonic moments with $E_l{\,>\,}\unit[1]{GeV}$.}
	\label{tab:LTreeCut}
\end{table}

Radiative corrections will give additional contributions proportional to $\alpha_s / \pi$ according to the expansion
\begin{equation}
L_n = L_{n,0} + \frac{\alpha_s}{\pi} L_{n,1} \qquad \mbox{and} \qquad
H_{ij} = H_{ij,0} + \frac{\alpha_s}{\pi} H_{ij,1}
\end{equation} 
which is again split into the various contributions of the different coupling constants as defined in (\ref{lskducnem}). Table~\ref{tab:LAlphas} contains the numerical results for the coefficients $L_n$, which we give according to the discussion of the last section only for the $c_L c_R$ contribution. The radiative corrections are sizable as expected from the ones for the total rate, but not abnormally large.

\begin{table}
	\begin{center}
	\begin{tabular}{c|d3d3}\toprule
		n & \multicolumn{1}{c}{$\:\ c_l^2$} & \multicolumn{1}{c}{$c_lc_r$} \\
		\midrule
		0 & -1.778 &  2.198 \\
		1 & -0.551 &  0.666 \\
		2 & -0.188 &  0.222 \\
		3 & -0.068 &  0.079 \\
		\bottomrule
	\end{tabular}
	$ \qquad \qquad$ 
	\begin{tabular}{c|d3d3}\toprule
		n & \multicolumn{1}{c}{$\:\ c_l^2$} & \multicolumn{1}{c}{$c_lc_r$} \\
		\midrule
		0 & -1.439 & 1.749 \\
		1 & -0.498 & 0.596 \\
		2 & -0.180 & 0.213 \\
		3 & -0.071 & 0.086 \\
		\bottomrule
	\end{tabular} 
    \end{center}
	\caption{$\alpha_s/\pi$ coefficients of the leptonic moments without $E_l$ cuts (left)
	and with a cut $E_l{\,>\,}\unit[1]{GeV}$ (right).}
	\label{tab:LAlphas}
\end{table}

The tree level result for the hadronic moments $H_{ij}$ is proportional to $\delta (M_{\rm had}^2-m_c^2)$ which means that the tree level contribution vanishes for $i>0$:
\begin{equation}
H_{ij,0} = 0 \,\, \mbox{for} \,\, i > 0.
\end{equation}
As discussed above, only the $\alpha_s$ corrections will yield a nontrivial shape of the spectrum. Thus we list in table~\ref{tab:HAlphas} the $\alpha_s/\pi$ coefficients of the hadronic moments without a lepton-energy cut, and in table~\ref{tab:HAlphasCut} the same quantity including a cut of $E_l > 1$ GeV.

\begin{table}
	\begin{center}
	\begin{tabular}{cc|d5d5}\toprule
		i & j & \multicolumn{1}{c}{$c_l^2$} & \multicolumn{1}{c}{$\:\ c_lc_r$} \\
		\midrule
		0 & 0 &  1.00000 & -0.66845 \\
		0 & 1 &  0.42200 & -0.25000 \\
		0 & 2 &  0.18319 & -0.09640 \\
		0 & 3 &  0.08147 & -0.03825 \\
		\bottomrule
	\end{tabular}
	$ \qquad \qquad$
	\begin{tabular}{cc|d4d4}\toprule
		i & j & \multicolumn{1}{c}{$c_l^2$} & \multicolumn{1}{c}{$\:\ c_lc_r$} \\
		\midrule
		0 & 0 &  0.8148 & -0.5617 \\
		0 & 1 &  0.3341 & -0.2037 \\
		0 & 2 &  0.1411 & -0.0761 \\
		0 & 3 &  0.0612 & -0.0293 \\
		\bottomrule
	\end{tabular}
	\caption{Tree level coefficients of the hadronic moments for $i=0$ without $E_l$ cuts (left) and with a cut  $E_l{\,>\,}\unit[1]{GeV}$ (right).}
	\label{tab:HTree}
	\end{center}
\end{table}

\begin{table}
	\begin{center}
	\begin{tabular}{cc|d3d3}\toprule
		i & j & \multicolumn{1}{c}{$c_l^2$} & \multicolumn{1}{c}{$c_lc_r$} \\
		\midrule
		0 & 0 & -1.778 &  2.198 \\
		0 & 1 & -0.719 &  0.867 \\
		0 & 2 & -0.292 &  0.349 \\
		0 & 3 & -0.118 &  0.143 \\
		\bottomrule
	\end{tabular}
	$ \qquad \qquad$
	\begin{tabular}{cc|d3d3}\toprule
		i & j & \multicolumn{1}{c}{$c_l^2$} & \multicolumn{1}{c}{$c_lc_r$} \\
		\midrule
		0 & 0 & -1.440 & 1.748 \\
		0 & 1 & -0.577 & 0.672 \\
		0 & 2 & -0.234 & 0.265 \\
		0 & 3 & -0.096 & 0.107 \\
		\bottomrule
	\end{tabular}
	\caption{$\alpha_s/\pi$ coefficients of the hadronic moments for $i=0$ without $E_l$ cuts (left) and with a cut  $E_l{\,>\,}\unit[1]{GeV}$ (right).}
	\label{tab:HAlphasI=0}
	\end{center}
\end{table}

\begin{table}
	\begin{center}
	\begin{tabular}{cc|d5d5d5d5d5d5}\toprule
		i & j & \multicolumn{1}{c}{$c_l^2$} & \multicolumn{1}{c}{$\:\ c_lc_r$} & \multicolumn{1}{c}{$c_ld_l$} & \multicolumn{1}{c}{$\:\ c_ld_r$} & \multicolumn{1}{c}{$c_lg_l$} & \multicolumn{1}{c}{$c_lg_r$} \\
		\midrule
		1 & 0 & 0.09009 & -0.03629 & 0.01697 & -0.05009 & 0.01286 & 0.06051 \\
		1 & 1 & 0.04700 & -0.01782 & 0.00789 & -0.02426 & 0.00679 & 0.03264 \\
		1 & 2 & 0.02509 & -0.00903 & 0.00377 & -0.01205 & 0.00364 & 0.01794 \\
		2 & 0 & 0.00911 & -0.00330 & 0.00117 & -0.00418 & 0.00121 & 0.00660 \\
		2 & 1 & 0.00534 & -0.00188 & 0.00062 & -0.00229 & 0.00071 & 0.00396 \\
		3 & 0 & 0.00181 & -0.00063 & 0.00018 & -0.00070 & 0.00023 & 0.00138 \\
		\bottomrule
	\end{tabular}
	\caption{$\alpha_s/\pi$ coefficients of the hadronic moments without $E_l$ cuts.}
	\label{tab:HAlphas}
	\end{center}
\end{table}

\begin{table}
	\begin{center}
	\begin{tabular}{cc|d5d5d5d5d5d5}\toprule
		i & j & \multicolumn{1}{c}{$c_l^2$} & \multicolumn{1}{c}{$\:\ c_lc_r$} & \multicolumn{1}{c}{$c_ld_l$} & \multicolumn{1}{c}{$\:\ c_ld_r$} & \multicolumn{1}{c}{$c_lg_l$} & \multicolumn{1}{c}{$c_lg_r$} \\
		\midrule
		1 & 0 & 0.05726 & -0.02300 & 0.00951 & -0.04187 & 0.00757 & 0.03344 \\
		1 & 1 & 0.02850 & -0.01069 & 0.00414 & -0.01980 & 0.00385 & 0.01727 \\
		1 & 2 & 0.01447 & -0.00510 & 0.00185 & -0.00958 & 0.00199 & 0.00905 \\
		2 & 0 & 0.00444 & -0.00157 & 0.00049 & -0.00302 & 0.00057 & 0.00274 \\
		2 & 1 & 0.00242 & -0.00082 & 0.00024 & -0.00158 & 0.00031 & 0.00153 \\
		3 & 0 & 0.00064 & -0.00021 & 0.00005 & -0.00042 & 0.00008 & 0.00041 \\
		\bottomrule
	\end{tabular}
	\caption{$\alpha_s/\pi$ coefficients of the hadronic moments with $E_l{\,>\,}\unit[1]{GeV}$.}
	\label{tab:HAlphasCut}
	\end{center}
\end{table}

The tables show that the various moments have a strong sensitivity to the non-standard couplings. With very few exceptions the coefficients of the new contributions are of the same order as the $c_L^2$ term, which contains the standard model. Thus if the moments can be determined at the level of a few percent this would result in a determination of the non-standard couplings at a similar level. However, a precise statement concerning the precision of an extraction of a non-standard model coupling is difficult, since a full analysis requires a combined fit of all the parameters, including the heavy quark parameters and heavy quark masses, which is beyond the scope of the present paper.

\section{Conclusions} 
In this paper we have computed the effect of a non-standard coupling for the $b \to c$ transition in a semileptonic decay, assuming that the leptonic current is purely left handed. The latter assumption is well justified by the data on purely leptonic processes, in particular the muon decay.

Although the calculation of nonperturbative effects for these non-standard couplings is straightforward, the main corrections are the perturbative QCD effects, which are sizable and have to be taken into account. Due to the vanishing anomalous dimension of the left and the right handed currents the QCD effects are finite for these currents; however, additional work is required to compute the virtual corrections to the scalar and tensor currents, which renormalize under QCD. Fortunately, these virtual corrections do not change the shape of the spectra and thus statements about moments are still possible.

On general grounds one would not expect new physics to show up in a charged current interaction, but this may as well be a false prejudice. At least from the generic point of view the most general parametrizations in terms of dimension six operators allow new physics effects in charged currents. As an example for a model one can consider a multi-Higgs Model, where a charged Higgs boson would induce an effect in a charged current already at tree level.

The effect of non-standard couplings on the moments can be sizable and thus this method opens a road to constrain possible new physics effects in charged current interactions. However, a detailed analysis needs a combined fit of all parameters in the heavy quark expansion, including the quark masses and the heavy quark expansion parameters, and the additional results given in this paper will allow us to perform such a fit.

\subsection*{Note added:}
When this paper was almost completed a parallel computation was communicated to us by I. Bigi with similar results. We thank I. Bigi and G. Song for communicating the results of their work prior to publication.

\subsection*{Acknowledgements}
We acknowledge useful discussions with O. Buchm\"uller and J. K\"uhn. Parts of the calculations were done with FeynArts \cite{FeynArts}, FormCalc and LoopTools \cite{FormCalcLoopTools}.
This work was partially supported by
the German Research Foundation (DFG) under contract No.
MA1187/10-1, and by the German Minister of Research (BMBF), contract No. 05HT6PSA.
\enlargethispage{1cm}

\end{document}